\documentclass{optica-article}
\journal{opticajournal} % for journals or Optica Open
\articletype{Research Article}
\usepackage{gensymb}
\usepackage{graphicx}
\usepackage{subcaption}
\usepackage{xcolor}

\begin{document}

%%%%%%%%%%%%%%%%%%%%%%%%%%    %%%%%%%%%%%%%%%%%%%%%%%%%%
\title{Virtually structured illumination for terahertz super-resolution imaging}

\author{James P. Fleming\authormark{1, 2, *}, Lucy A. Downes\authormark{2}, John M. Girkin\authormark{1} and Kevin J. Weatherill\authormark{2}}

\address{\authormark{1}Centre for Advanced Instrumentation, Department of Physics, Durham University,
South Road, Durham DH1 3LE, United Kingdom\\

\authormark{2}Quantum Light and Matter Group, Department of Physics, Durham University,
South Road, Durham DH1 3LE, United Kingdom}

\email{\authormark{*}james.p.fleming@durham.ac.uk}

%%%%%%%%%%%%%%%%%%%%%%%%%%  ABSTRACT  %%%%%%%%%%%%%%%%%%%%%%%%%%
% use {asbstract*} to suppress the copyright line. Copyright information will be added in production
\begin{abstract*} 
We demonstrate structured illumination super-resolution imaging in the Terahertz (THz) frequency band using the Virtually Structured Detection (VSD) method. Leveraging our previously reported high-speed, high-sensitivity atomic-based THz imager, we achieve a resolution enhancement of $(74\pm3)\%$ at 0.55 THz, without the aid of deconvolution methods. We show a high-speed THz imaging system is compatible with the use of advanced optical techniques, with potential disruptive effects on applications requiring both high speed and high spatial resolution imaging in the THz range.
\end{abstract*}

%%%%%%%%%%%%%%%%%%%%%%%%%%  BODY  %%%%%%%%%%%%%%%%%%%%%%%%%%
\section{Introduction}
% - Highlighting THz imaging applications
Imaging in the THz band (0.3-3~THz) \cite{mittleman2018review} has many applications ranging from security screening \cite{tonouchi2007cutting} and non-destructive testing \cite{brinkmann2017towards} to biomedical diagnostics \cite{yang2016biomedical}. These applications take advantage of the favorable properties of the THz frequency range, such as its ability to penetrate common non-conducting materials like wood, paper, and plastics \cite{leitenstorfer2023}. Additionally, its non-ionizing nature makes it a safer alternative to higher-frequency penetrating radiation, further broadening its appeal across various domains. 

% - Two THz techniques - SP and FPA
THz imaging can broadly be categorized into two general schemes: single-point (SP) detection and focal-plane array (FPA) detection. While SP detectors typically acquire images by raster-scanning a target object, FPA detectors can capture images in a single shot using a 2D array of sensors for parallel acquisition. Through the use of far-field aperture-based raster scanning techniques, SP detectors can achieve sub-diffraction limited spatial resolution and, when combined with time-domain spectroscopy (TDS), can produce images with hyperspectral information. However, the raster-scanning process is inherently slow, making SP detection unsuitable for applications requiring rapid image acquisition. % - FPAs have the advantage of speed but not resolution
% - Recent improvement of room-temp 'THz Cameras' leading to near-video rate acquisition 
In contrast, FPA detectors offer significantly faster imaging speeds by capturing images in a single-shot. Recent advancements in room-temperature FPAs have led to commercially available THz cameras with video-rate acquisition performance. However, despite these advances, THz FPA detectors generally suffer from low sensitivity, high noise, provide no spectral information, and ultimately are constrained by diffraction-limited spatial resolution. 

% - Techniques in optical microscopy have been developed for surpassing the diffraction-limit. SIM is a widefield technique that offers twice the diffraction-limit resolution
Techniques to overcome the diffraction limit have been investigated in optical microscopy for many years. One example of such super-resolution techniques is Structured Illumination Microscopy (SIM) \cite{gustafsson2000sim, heintzmann2009subdiffraction}. SIM involves illuminating a sample with a known spatial pattern to allow the extraction of high-frequency spatial information. This technique can  increase spatial resolution of images by up to a factor of two above the diffraction limit \cite{chen2023sim-review}. One downside of SIM is that it requires the acquisition of numerous images in order to collect sufficient spatial information. Whilst this is not a major issue for optical imaging where sensors are fast and have low noise, it is a more significant problem in the THz range and previous attempts of super-resolution imaging at THz frequencies have either been purely through computational enhancement \cite{long2019terahertz, wang2021terahertz} or demonstrated via SP detection \cite{guerboukha2020super}.

% - Novel highly sensitive imaging allows for ultra-fast imaging
Recently, a novel THz FPA sensor based on THz-to-optical conversion in room temperature, optically pumped atomic vapor has emerged~\cite{wade17,lucy2020kilohertz}. This technique has demonstrated very high image acquisition rates in comparison to other room-temperature FPA sensors, and because images are captured using standard optical cameras, sensors can be chosen to match the application.

Here, we make use of a fast atom-based THz imaging system to apply a variant of SIM known as Virtually Structured Detection (VSD) \cite{lu2013vsd, lu2021vsd} to achieve THz super-resolution imaging. Being able to perform super-resolution imaging at 0.55~THz holds promise for applications that require high-speed and high-spatial resolution, while leveraging the deep penetrating properties of 0.55~THz radiation.

\section{Virtually Structured Detection}
SIM often relies on complex illumination systems that are impractical at THz frequencies \cite{wei2014psim ,dan2013dmdsim, Fiolka2008slm}. VSD simplifies this by using a point or line illumination to scan the object, capturing a full-field image at each scan position. For THz VSD, we employ a scanning line illumination \cite{lu2013vsd, lu2021vsd, zhi2015vsd}, enabling low-complexity THz super-resolution 
imaging. VSD reconstruction begins by taking each image scan, multiplying it by a modulation mask $m(\mathbf{r})$, and spatially integrating it along the scan axis - reducing it to a single-pixel-wide image \cite{lu2013vsd}. This processed scan $p_i(\mathbf{r_i})$ is represented as

\begin{equation}
    p_i(\mathbf{r_i}) = h_{il}(\mathbf{r}) \otimes \left\{ \left[ h_{de}(\mathbf{r}) \otimes m(\mathbf{r})  \right] s(\mathbf{r})\right\}~,
\end{equation}
where $\mathbf{r} \equiv (x, y)$ is the spatial position vector, $s(\mathbf{r})$ the sample object, $h_{il}(\mathbf{r})$ and $h_{de}(\mathbf{r})$ are the point spread functions (PSF) of the respective illumination and detection paths of the imaging system, and $\otimes$ represents the image convolution operator. Successive scans are then stacked to generate a virtually structured image $p(\mathbf{r}) = \left[ p_1(\mathbf{r_1}) \, \cdots \, p_n(\mathbf{r_n}) \right]$, which is mathematically equivalent to those acquired in conventional widefield SIM \cite{zhi2015vsd}. As with widefield SIM, a sinusoidal modulation $m(\mathbf{r}) = \cos\left(2\pi \mathbf{p_\theta} \cdot \mathbf{r} + \phi\right)$ is employed, where $\mathbf{p_\theta}\equiv(\mathrm{p}\cos\theta,~\mathrm{p}\sin\theta)$ with $\theta$ being the scan angle and $\phi$ the modulations phase. The significance of such modulation becomes clearer from its Fourier transform, $\tilde{M}(\mathbf{k})$:

\begin{equation}
    \tilde{M}(\mathbf{k}) = \delta\left(\mathbf{k} + \mathbf{p_\theta}\right)e^{i\phi} +
                    \delta\left(\mathbf{k} - \mathbf{p_\theta}\right)e^{-i\phi}~.
\end{equation}

Here, the modulation comprises of two delta functions, $\delta\left(\mathbf{k} \pm \mathbf{p_\theta}\right)$, which when convolved with the Fourier spectrum of the object, $\tilde{S}(\mathbf{k})$, shifts the content of the spectrum by the spatial frequency vector, $\pm \mathbf{p_\theta}$. This is a feature of the Fourier shifting property \cite{adams2019f2f}. Under such modulation $\tilde{P}(\mathbf{k})$, the Fourier transform of $p(\mathbf{r})$, is realized as

\begin{equation}
    \tilde{P}(\mathbf{k}) = 
        \tilde{H}(\mathbf{k})
        \left[ 
            \tilde{S}(\mathbf{k} + \mathbf{p_\theta})e^{i\phi} +
            \tilde{S}(\mathbf{k} - \mathbf{p_\theta})e^{-i\phi}
        \right]~.
    \label{eq:fourier_picture}
\end{equation}

These shifted Fourier bands $\tilde{S}(\mathbf{k} \pm \mathbf{p_\theta})$ contain higher spatial frequency information that otherwise would be cut-off by the optical transfer function (OTF) of the imaging system $\tilde{H}(\mathbf{k})= \tilde{H}_{il}(\mathbf{k}) \tilde{H}_{de}(\mathbf{k})$. Solving Eq.~\eqref{eq:fourier_picture} for the shifted Fourier bands $\tilde{S}(\mathbf{k}\pm\mathbf{p}_\theta)$ by changing the phase of the modulation, $\phi$, allows for retrieval of the two shifted bands,

\begin{equation}
    \tilde{H}(\mathbf{k}) \tilde{S}(\mathbf{k} \pm \mathbf{p_\theta}) 
        = \tilde{P}(\mathbf{k}, \phi=0) \pm i\tilde{P}(\mathbf{k}, \phi=\pi/2)~.
        \label{eq:recovered_bands}
\end{equation}

As the modulation is applied computationally, no additional image scans are required for the additional phase. Unlike SIM, the reconstruction is phase artifact-free as the phase is exactly defined \cite{zhi2015vsd}. Typically, attenuation of spatial frequencies caused by $\tilde{H}(\mathbf{k})$ can be restored via a final deconvolution step

\begin{equation}
   \tilde{S}(\mathbf{k} \pm \mathbf{p_\theta}) 
        = \text{deconvolve}\left\{\tilde{P}(\mathbf{k}, \phi=0) \pm i\tilde{P}(\mathbf{k}, \phi=\pi/2)\right\}~.
\end{equation}

For this work, we opted not to deconvolve our images to assess the standalone performance of the VSD technique. We direct the reader to the extensive use of various deconvolution algorithms in the literature \cite{muller2016fairSIM, ingaramo2014deconvRL, perez2016deonvRL, chu2014deconvTV}.

Super-resolution image reconstruction concludes by summation of the two shifted Fourier bands, improving the spatial frequency extent for a theoretical maximum two-fold resolution when $|\mathbf{p_\theta}|=\mathrm{k_{cutoff}}$, the cut-off spatial frequency of the widefield system \cite{lal2016simalgo}. Such improvement is observed across the scan-axis only. For near-isotropic resolution improvement, the angle over which the sample is scanned is varied, typically as $\theta=0\degree, 60\degree$ and  $120\degree$.

\section{Experimental Implementation}
\begin{figure}[ht]
\centering\includegraphics[height=5.9cm]{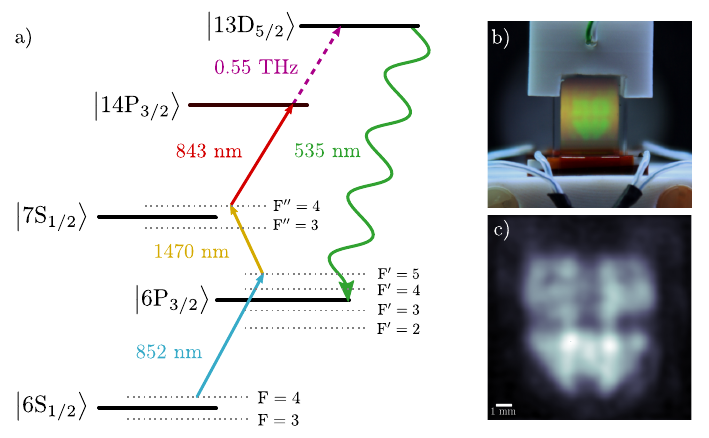}
\caption{a) Energy ladder scheme in Cesium used for the conversion of $0.55~\mathrm{THz}$ to the green $535~\mathrm{nm}$ via decay fluorescence. b) True-color composite image of the vapor cell with $1~\mathrm{cm}^2$ sensing area, both unfiltered background fluorescence and green signal fluorescence are shown. c) Processed monochromatic THz image from the acquired fluorescence signal.}
\label{fig:ladder_scheme}
\end{figure}

Widefield THz images were collected using our atom-based imaging system, full details of which are available in previous publications \cite{lucy2020kilohertz, lucy2023tutorial}. To summarize, using infrared lasers, frequency stabilized to atomic references \cite{polspec,brynpaper}, we can promote alkali metals to highly excited states, known as Rydberg states \cite{gallagher1994rydberg}. Rydberg states feature electric-dipole transitions that fall within the THz frequency band and are therefore sensitive to resonant THz fields. Rydberg states also spontaneously decay via probabilistic decay pathways back to the atomic ground state, producing an optical fluorescence spectra. By illuminating a thermal vapor of optically pumped alkali metal atoms with a resonant THz-field, the appropriate Rydberg transition can be driven, which will then decay to produce a detectable optical fluorescence signal. By shaping each laser with a combination of cylindrical lenses, overlapping lightsheets can be formed, constructing an active sensing focal-plane. High-speed, high-sensitivity focal-plane THz imaging is therefore realized by efficient THz-to-optical conversion. The sub-microsecond decay rates of the Rydberg states allows theoretical imaging rates in the 1 MHz range. In practice the imaging frame-rate is limited by the signal-to-noise and available throughput of the optical readout system, which has been demonstrated to a maximum of 12,000 frames per second, an almost three-orders of magnitude improvement when compared to the state-of-the-art commercial THz cameras \cite{lucy2023tutorial}.

In this study, we operate our Rydberg-based THz imaging system at $0.55~\mathrm{THz}$ by using the 14P$_{3/2} \rightarrow$ 13D$_{5/2}$ transition in Cesium, as shown in Fig.~\ref{fig:ladder_scheme}a. A Cesium thermal vapor (Fig.~ \ref{fig:ladder_scheme}b) is contained in a heated quartz cell and maintained at 45$^\circ$, which maximizes the fluorescence signal without the thermal vapor becoming optically thick to the infrared excitation lasers. The vapor cell provides optical access on all four sides. Three frequency-stablized infrared lasers ($\lambda=852~\mathrm{nm},~1470~\mathrm{nm},~843~\mathrm{nm}$) drive the three-step excitation scheme and prepare Cesium in the 14P$_{3/2}$ state. The size of the vapor cell and available laser power produces an active sensing region of $1~\mathrm{cm}^2$. Narrow-band, resonant $550~\mathrm{GHz}$ continuous-wave THz illumination is provided by a Virginia Diodes Inc. amplifier multiplier chain (AMC). A stable microwave generator seeds the AMC, and the frequency is upconverted by a factor of 36, providing a maximum power output of $5~\mathrm{mW}$. A power output of $0.3~\mathrm{mW}$ was used to illuminate the target without saturating the imaging system.

\begin{figure}[tbp]
\centering\includegraphics[width=\textwidth]{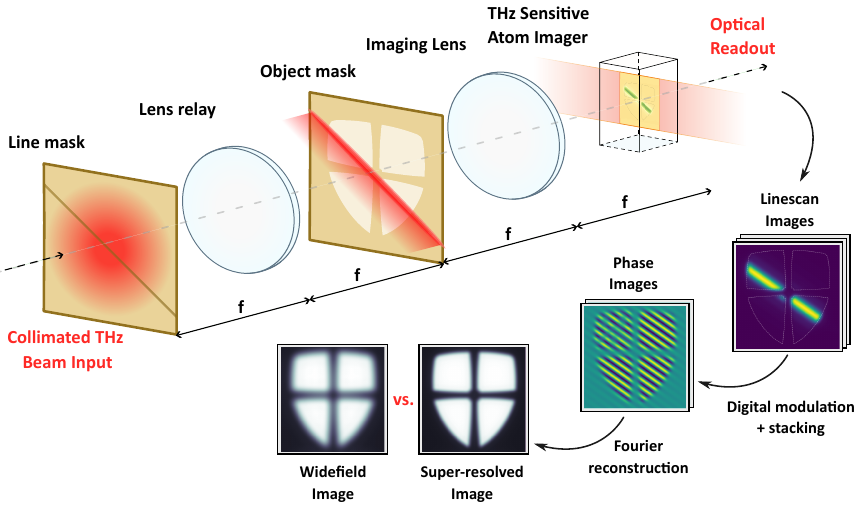}
\caption{
Diagrammatic overview of the THz VSD implementation, showing the working elements required to generate full-frame image scans. Our atomic imager facilitates efficient THz-to-optical conversion for fast, full-frame optical readout via a monochromatic CCD (not shown). To generate a virtually structured image, successive images are taken as the object is translated through the THz line illumination. These are digitally modulated, spatially integrated and stacked to produce the phase images, which are used to reconstruct the super-resolve image. }
\label{fig:experimental_layout}
\end{figure}

To implement VSD, the required line-profile illumination was generated via a mask with a sub-diffraction, 250~$\mathrm{\mu m}$ wide slit. The illuminated mask was then imaged onto the object plane using an image relay of two aspheric PTFE plano-convex lenses $(\mathrm{d}=50.8~\mathrm{mm}, \mathrm{f}=75~\mathrm{mm})$, as shown in Fig.~\ref{fig:experimental_layout}. The mask was mounted on a manual rotation stage in order to set the scan angle $\theta$. Unlike previous optical implementations\cite{lu2013vsd, lu2021vsd, zhi2015vsd}, the scanning process occurs by translating the object normal to the fixed line illumination, using a pair of translation stages (Thorlabs MTS50-Z8). A transmissive mask was used as the object of interest to image, placed at the output plane of the image relay.  Both illumination and object masks were produced by milling copper-clad FR4 board. A common material used in electronic circuit board production, the copper foil cladding blocks transmission, while FR4 substrate is transmissive at 0.55 THz.

Widefield full-frame images of the object are taken using our atomic THz imager. A commercial 1x silicon objective lens $(\mathrm{i2S}, \mathrm{d}=60~\mathrm{mm}, \mathrm{f}=70~\mathrm{mm})$ images onto the THz-sensitive atomic vapor. The optical fluorescence from the vapor was captured by an Andor iXon Ultra 888 EMCCD camera, under a 200~ms exposure. An appropriate bandpass filter was used to isolate the signal from superfluous background fluorescence. Images are post-processed by subtracting a background frame to further isolate the signal \cite{lucy2023tutorial}. Each captured image was then cropped to a region of interest (ROI) around the line-center of the illumination. A ROI with width of the expected Airy disc size was found to be an adequate trade-off between signal-to-noise and rejection of unwanted diffracted illumination. To generate comparative true widefield images, the illumination mask was removed and the object mask centered to produce an equivalent uniformly illuminated, single-shot widefield image.

\section{Results}

\begin{figure}[htbp]
    \centering\includegraphics[width=\textwidth]{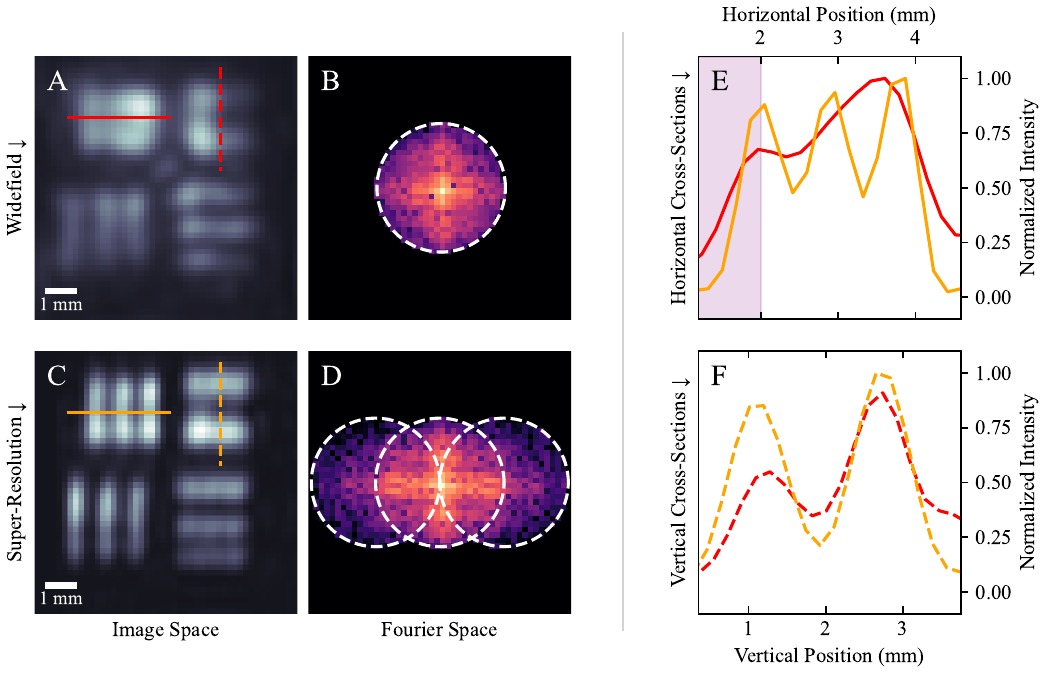}
    \caption{Widefield (A) and super-resolution (C) images of a USAF resolution target with their respective Fourier spectra (B and D). Cross-sections of the vertical bars (E) of Element 2 highlight resolution improvement along the horizontal scan direction only. The indicated purple region marks the edge used for quantitative analysis resolution improvement.}
    \label{fig:resolution_target}
\end{figure}

To validate the VSD method, Elements 1 and 2 of Group 0 on a USAF resolution target were imaged in both the widefield and VSD modality. The imaged elements correspond to a lines-per-millimeter spacing of $1.00~\mathrm{lp/mm}$ and $1.12~\mathrm{lp/mm}$ respectively. For VSD reconstruction, the target was scanned along the horizontal direction, to produce an improved resolution along the horizontal only. A total of 48 scan images were taken, leading to a reconstructed super-resolution image size of $48\times48$ pixels. This size was chosen such that the super-resolution image meets the Nyquist sampling limit without oversampling \cite{ruzin2024techniques}. Fig.~\ref{fig:resolution_target} presents a comparison of widefield (A) and super-resolution (C) imaging results, including cross-sectional profiles of the vertical (E) and horizontal (F) bars of Element 2. The cross-section of the vertical bars (E) shows successful super-resolution, with the vertical bars being well resolved in the super-resolution image (C). In contrast, the widefield image (A) fails to resolve the vertical bars, as their spatial frequency exceeds the calculated cut-off of the imaging lens. In contrast, the cross-section of the horizontal bars (F) for the super-resolution and widefield image show the horizontal bars are unresolved in both. This is consistent with the resolution enhancement being limited to the horizontal axis. The widefield Fourier spectrum (B) and super-resolution Fourier spectrum (D) is also shown, with contours showing the extent of each Fourier band. The extent of the central Fourier bands in both images is determined by the spatial cut-off frequency of the imaging system. For the super-resolution Fourier spectrum, two additional recovered Fourier bands (\autoref{eq:recovered_bands}) improve the extent of Fourier spectra along the horizontal direction. This improved extent results in the the improvement in image resolution. 

To quantify the improvement in resolution, the edge profile of one of the horizontal bars in Element 2 was analyzed. This is highlighted by the purple shaded region in (E). This edge is representative of the Edge Spread Function (ESF) of the imaging system, which is the cumulative distribution of the system's Point Spread Function (PSF) \cite{barakat1965determination}. We characterize the ESF of both the widefield and super-resolution image by fitting the Gaussian error function $\mathrm{erf}$, and taking the derivative to produce a Gaussian approximation of the PSF. This resulted in a PSF full width half maximum of $0.94\pm0.01~\textrm{mm}$ for the widefield image and $0.54\pm0.01~\textrm{mm}$ for the super-resolution image. This represents a $(74\pm3)\%$ improvement in resolution. This resolution enhancement is lower than the theoretically possible factor of two and is limited by the presence of image noise and attenuation of high spatial frequencies in the Fourier domain. Both factors, however, can be corrected for through the omitted deconvolution step \cite{lal2016simalgo}.

\begin{figure}[htbp]
    \centering
    \begin{subfigure}[b]{0.45\textwidth}
        \includegraphics[width=\textwidth]{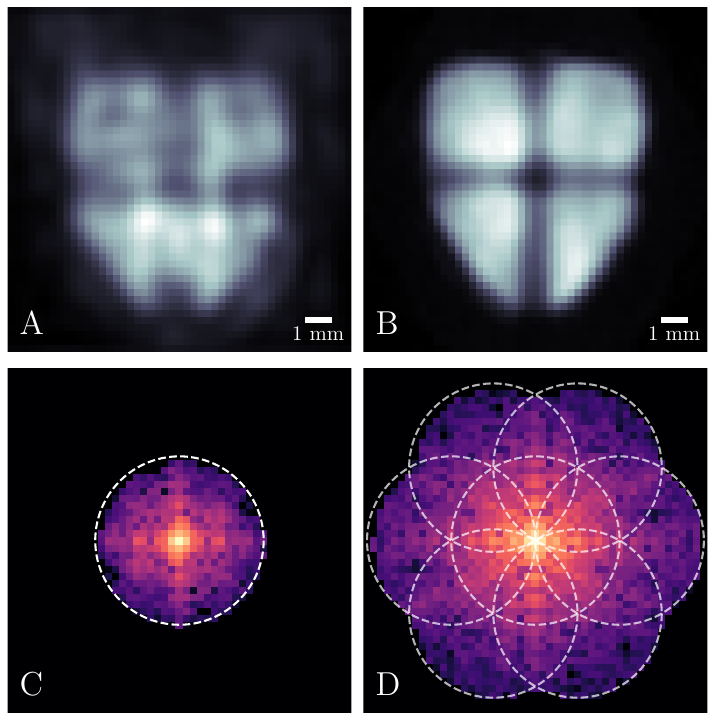}
        \label{fig:plot1}
        \caption{}
    \end{subfigure}
    \hspace{0.025\textwidth} % Horizontal space between figures
    \begin{subfigure}[b]{0.45\textwidth}
        \includegraphics[width=\textwidth]{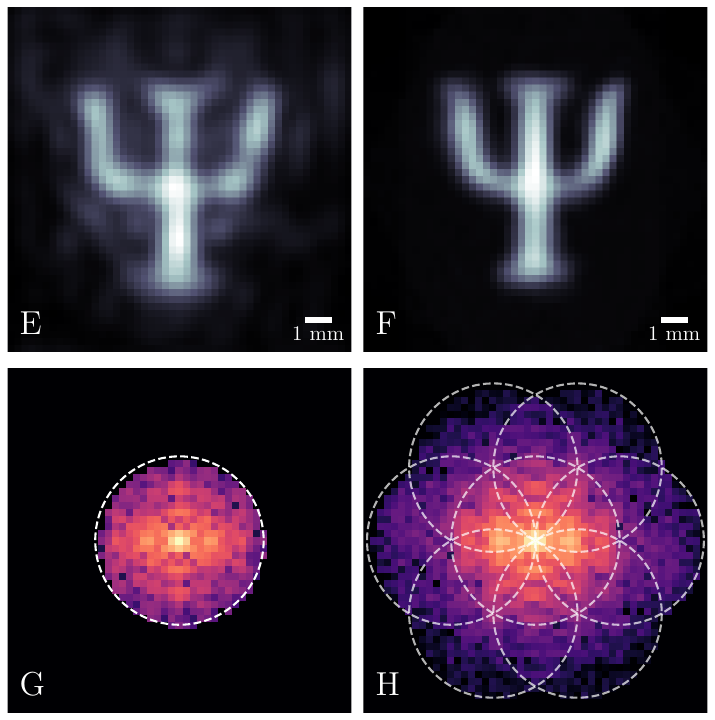}
        \label{fig:plot2}
        \caption{}
    \end{subfigure}
    
    % Shared caption for all subplot
    \caption{Comparison of true wide-field images (A, E) and super-resolution images (B, F) of two transmission targets a) the Durham University shield b) the Greek letter $\Psi$. Each image is accompanied by its respective Fourier spectrum, demonstrating an increased spatial frequency content in the super-resolution images. }
    \label{fig:isotropic_super_resolution}
\end{figure}

To show VSD can be used to produce isotropic resolution improvement, two pictographic targets were imaged; the Durham University shield and the Greek letter $\Psi$. Both image icons measured $8.5\times9.7~\mathrm{mm}$ to maximize the coverage across the available field-of-view of the THz imager. The respective widefield and super-resolution images are shown in Fig.~\ref{fig:isotropic_super_resolution} A, B, E and F. The qualitative image masks were scanned along three equally spaced axes at angle $\theta=0\degree, 60\degree, 120\degree$, this was achieved by rotating the illumination mask and translating the image mask normal to the illumination. Each scan axis comprised of a total of 48 scan images and provides resolution improvement along its axis. Therefore 144 full-frame images were required to produce one super-resolution image of a $10\times10~\mathrm{mm}$ field-of-view. In the Fourier domain, the super-resolution images are constructed from seven distinct Fourier bands; two shifted bands per scan axis, plus one central band (Fig.~\ref{fig:isotropic_super_resolution} D, H). To improve signal-to-noise, the central band is the mean average of the recovered central bands from the three scans.  The seven distinct Fourier bands are combined to produce near-isotropic extension of the Fourier spectra. For all Fourier domain images (Fig.~\ref{fig:isotropic_super_resolution} C, D, G, H), each Fourier band is highlighted by contours, and as with Fig.~\ref{fig:resolution_target} the extent of contour is determined by the calculated spatial cut-off frequency of the imaging system. By comparison, the super-resolution images exhibit better resolved detail and high image contrast along edges, this is due to the improved extent of the respective Fourier spectra from the VSD reconstruction process. 

\section{Conclusion}

This work has demonstrated experimental implementation of structured illumination super-resolution imaging in the THz band using virtually structured detection, producing a quantitative resolution improvement of $(74\pm3)\%$. While less than the theoretical maximum of two-fold resolution improvement, this has been demonstrated without the typical deconvolution step that aides recovery of higher spatial frequencies. While SIM-inspired imaging has been considered in the neighboring microwave band \cite{Shayei2018mwsim}, to our knowledge, this is the longest wavelength at which such technique has been demonstrated experimentally.

A low complexity implementation was prioritized over imaging acquisition speed, in order to evaluate the VSD technique.  While the VSD method is inherently slower than widefield SIM due to the requirement of more acquired images per reconstruction, the high-speed nature of our atomic THz imager meant overall acquisition speed was limited by the translation rate of the object. A faster acquisition scheme could be achieved using a beam-scanning system, vastly increasing the acquisition rates at the expense of increased operational complexity.  This would offer video-rate THz imaging at high spatial resolution while utilizing the advantageous penetrating properties of $0.55~\mathrm{THz}$ through materials, such as FR4 circuit board, that are opaque at higher THz frequencies.

\section{Acknowledgements}

We thank Andrew MacKellar for stimulating discussions about both experimental and data processing aspects of this work. This work is supporrted by Engineering and Physical Sciences Research Council (EPSRC) grants EP/W033054/1 and EP/S015973/1. JPF is supported by EPSRC studentships.

\section{Disclosures}
The authors declare no conflicts of interest

\section{Data availability}
Data underlying the results presented may be obtained from the authors upon reason-
able request.

\bibliography{sample}
\end{document}